\documentclass[twocolumn,pra,showpacs,floatfix]{revtex4}
\usepackage{graphicx}                
\usepackage{dcolumn}                 
\usepackage{times}
\usepackage{amsmath}

\newcommand*{\cent}[1]{\multicolumn{1}{c}{$#1$}}
\newcolumntype{w}[1]{D{.}{.}{#1}}
\newcommand{\br}{\vec{r}}
\newcommand{\bs}{\vec{s}}
\begin{document}

\title{Electric dipole rovibrational transitions in HD molecule}

\author{Krzysztof Pachucki}
\email{krp@fuw.edu.pl}
\affiliation{Institute of Theoretical Physics,
             University of Warsaw, Ho{\.z}a 69, 00-681~Warsaw, Poland }
\author{Jacek Komasa}
\email{komasa@man.poznan.pl}
\affiliation{Faculty of Chemistry,
        A.~Mickiewicz University, Grunwaldzka 6, 60-780~Pozna\'n, Poland }

\date{\today}

\begin{abstract}
The rovibrational electric dipole transitions in the ground electronic state
of the HD molecule are studied. A simple, yet rigorous formula is derived for the 
transition rates in terms of the electric dipole moment function $D(R)$, which is 
calculated in a wide range of $R$. Our numerical results for transition rates are 
in moderate agreement with experiments and previous calculations, but are at 
least an order of magnitude more accurate.
\end{abstract}

\pacs{33.70.Ca, 95.30.Ky, 31.15.-p, 31.30.-i}
\maketitle

\section{Introduction}

The electric dipole rovibrational transitions in the HD molecule are possible due 
to different masses of the proton and of the deuteron and thus slightly different
binding energies in hydrogen and deuterium.  These transitions,
for the first time, were observed by Herzberg \cite{herz}
and since then measured by several groups
\cite{trefler, nelson,NT83, bejar, kellar1,RJM82, kellar4, dalby}.
Theoretical calculations of the dipole transition moment
were first carried
out by Wick \cite{wick}, somewhat later by Wu \cite{wu}
and by Blinder \cite{blinder}.
More elaborate calculations include those of Bunker \cite{bunker},
Wolniewicz \cite{wol}, Ford and Browne \cite{ford},
and Thorson {\em et al.} \cite{thorson}.
The most recent works \cite{wol,ford,thorson} are in
generally good agreement with experimental results in
\cite{nelson,NT83, bejar, kellar1,RJM82, kellar4, dalby}.

In this work we derive a compact formula for the dipole transition moment
using a unitary transformation of the Hamiltonian followed by the adiabatic 
approximation, and present results in terms of the electric dipole moment 
function $D(R)$. We obtain $D(R)$ for a wide range of internuclear distances 
$R\in\langle 0.5,12\rangle$ au, which enables calculations of electric dipole 
transitions between all rovibrational states including the highly excited ones. 
Although they have not been measured, these dipole transitions
between highly excited states
together with electric quadrupole transitions lead to the cooling 
of the hydrogen clouds in the interstellar space \cite{abgrall}. 
The obtained transition rates between low lying rovibrational states
are the most accurate to date, and agree with experimental values
with minor exceptions.

\section{Derivation of the transition dipole moment}
In order to derive a formula for the dipole transition moment,
we consider a diatomic molecule in the reference frame of 
the geometrical center of the two nuclei. The total wave function $\phi$ is 
a solution of the stationary Schr\"odinger equation
\begin{equation}
H\,\phi = E\,\phi\,, \label{01}
\end{equation}
with the Hamiltonian
\begin{equation}
H = H_{\rm el} + H_{\rm n}\,, \label{02}
\end{equation}
split into the electronic and nuclear parts.  In the electronic Hamiltonian
\begin{equation}
H_{\rm el} = -\sum_{a}\frac{\nabla^2_a}{2\,m_{\rm e}} + V\,, \label{03}
\end{equation}
with $V$ including the Coulomb interaction,
the nuclei have fixed positions $\vec R_A$ (proton) and $\vec R_B$ (deuteron),
while the nuclear Hamiltonian is
\begin{eqnarray}
H_{\rm n}&=&- \frac{\nabla^2_{\!R}}{2\,\mu_{\rm n}}
           - \frac{\bigl(\sum_{a}\,\vec\nabla_a\bigr)^2}{8\,\mu_{\rm n}}
           \nonumber \\ &&
           - \frac{1}{2}\,\biggl(\frac{1}{M_B}-\frac{1}{M_A}\biggr)\,
             \vec\nabla_R\cdot\sum_a\vec\nabla_a\,,\label{04}
\end{eqnarray}
where $\vec R = \vec R_A-\vec R_B$ and $\mu_{\rm n}$ is the nuclear reduced
mass. In the adiabatic approximation the total wave function of the molecule
\begin{equation}
\phi_{\rm a}(\vec r,\vec R) = \phi_{\rm el}(\vec r)_{\!\vec R} \; \chi(\vec R) \label{05}
\end{equation}
is represented as a product of the electronic wave function $\phi_{\rm el}$
and the nuclear wave function $\chi$.  The electronic wave function obeys
the clamped nuclei electronic Schr\"odinger equation
\begin{equation}
\bigl[H_{\rm el}-E_{\rm el}(R)\bigr]\,|\phi_{\rm el}\rangle = 0, \label{06}
\end{equation}
while the wave function $\chi$ is a solution to the nuclear Schr\"odinger 
equation with the effective potential generated by electrons
\begin{equation}
\biggl[-\frac{\nabla_R^2}{2\,\mu_{\rm n}} +\bigl\langle\phi_{\rm el}| 
H_{\rm n}|\phi_{\rm el}\bigr\rangle +E_{\rm el}(R)-E_{\rm a}\biggr]\,
|\chi\rangle = 0\,, \label{07}
\end{equation}
where the so called diagonal (or adiabatic) correction
\begin{eqnarray}\label{Ead}
\langle\phi_{\mathrm{el}}|H_{\rm n}|\phi_{\mathrm{el}}\rangle&=&
\frac{1}{2\mu_{\mathrm{n}}}\langle\vec\nabla_R\phi_{\mathrm{el}}|
\vec\nabla_R\phi_{\mathrm{el}}\rangle \nonumber \\
&&-\frac{1}{8\mu_{\mathrm{n}}}\langle\phi_{\mathrm{el}}|
\Bigl(\sideset{}{_a}\sum\!\vec\nabla_a\Bigr)^2|
\phi_{\mathrm{el}}\rangle\,
\end{eqnarray}
is a function of $R$.

The existence of the electric dipole transitions in HD is due to
the last term in Eq.~(\ref{04}). This term can be used directly as 
a perturbation. Such an approach is presented
in the Appendix for a comparison with previous works. In an alternative
method, inspired by the work of Thorson {\em et al.}~\cite{thorson} 
and applied here, we introduce a unitary transformation 
\begin{equation}
H' = U^+\,H\,U \label{09}
\end{equation}
to shift the odd term in $H_{\rm n}$ to the potential $V$ in Eq.~(\ref{03}).
This transformation greatly simplifies further calculations.
We choose $U$ of the form
\begin{equation}
U = e^{\lambda\,(\vec r_1+\vec r_2)\cdot\vec\nabla_R}\label{10}
\end{equation}
with
\begin{equation}\label{11}
\lambda=-\frac{m_{\mathrm{e}}}{2}\,\biggl(\frac{1}{M_B}-\frac{1}{M_A}\biggr)
\end{equation}
and obtain $H'$ with neglecting $O\!\left[({m_{\rm e}}/{\mu_{\rm n}})^2\right]$ terms, namely
\begin{eqnarray}
H' &=& H + \lambda\,[H,(\vec r_1+\vec r_2)\cdot\vec\nabla_R]+ O(\lambda^2)
\nonumber \\
&=& H -\lambda\,(\vec r_1+\vec r_2)\cdot\vec\nabla_R(V)
-\frac{\lambda}{m_{\mathrm{e}}}\,(\vec\nabla_1+\vec\nabla_2)\cdot\vec\nabla_R\nonumber\\
&=& H_{\rm el} +\delta V + H'_{\rm n}\,, \label{12}
\end{eqnarray}
where 
\begin{eqnarray}
\delta V &=& \frac{m_{\rm e}}{2}\,\biggl(\frac{1}{M_B}-\frac{1}{M_A}\biggr)\,
(\vec r_1+\vec r_2)\cdot\vec\nabla_R(V)\label{13}\\
\vec\nabla_R(V) &=& \frac{1}{2}\,\biggl(-\frac{\vec r_{1A}}{r_{1A}^3}+\frac{\vec r_{1B}}{r_{1B}^3}
-\frac{\vec r_{2A}}{r_{2A}^3}+\frac{\vec r_{2B}}{r_{2B}^3}\biggr)-\frac{\vec n}{R^2}\quad\label{14}\\
H'_{\rm n} &=& - \frac{\nabla^2_{\!R}}{2\,\mu_{\rm n}}
           - \frac{\bigl(\sum_{a}\,\vec\nabla_a\bigr)^2}{8\,\mu_{\rm n}}\label{EHprime}
\end{eqnarray}
and  $\vec n = \vec R/R$.

The E1 transition between rovibrational levels
of the HD molecule in the ground electronic state comes now from the nonadiabatic
correction $\delta V$ to the electronic potential $V$.
In the leading order one uses the adiabatic approximation, and
the electric dipole moment $\vec D_{\rm fi}$ between some
initial $\phi_{\rm i}$ and final state $\phi_{\rm f}$ is
\begin{eqnarray}
\vec D_{\rm fi} &=& \langle\phi_{\rm f}|\vec r|\phi_{\rm i}\rangle\label{16}\\
            &=&
\langle\phi_{\rm el}\,\chi_{\rm f}|\vec r\,
\frac{1}{E_{\rm el}-H_{\rm el}}\,\delta V|\phi_{\rm el}\,\chi_{\rm i}\rangle
\nonumber \\ &&
+\langle\phi_{\rm el}\,\chi_{\rm f}|\delta V\,
\frac{1}{E_{\rm el}-H_{\rm el}}\,\vec r|\phi_{\rm el}\,\chi_{\rm i}\rangle\,,
\label{17}
\end{eqnarray}
where $\vec r = \sum_a \vec r_a$. We claim, without presenting the proof,
that the higher order nonadiabatic corrections are smaller by a factor of
$m_{\rm e}/\mu_n\approx 10^{-3}$, and their contribution to $\vec D_{\rm fi}$ 
can be neglected.

Below, we rewrite this matrix element in terms of the electric
dipole moment function $D(R)$, namely
\begin{eqnarray}
\vec D_{\rm fi} &=& \langle\chi_{\rm f}|D\,\vec n|\chi_{\rm i}\rangle
= \langle J_{\rm f},M_{\rm f}|\vec n|J_{\rm i},M_{\rm i}\rangle\,D_{\rm fi},\label{18}\\
D_{\rm fi} &=& \int dR\,R^2\,D(R)\,\chi^*_{J_{\rm f}}(R)\,\chi_{J_{\rm i}}(R),\label{19}\\
D(R) &\equiv& \biggl(\frac{m_{\rm e}}{M_B}-\frac{m_{\rm e}}{M_A}\biggr)\label{dr}
\\ &&\times
\langle\phi_{\rm el}|\vec r\cdot\vec n\,
\frac{1}{E_{\rm el}-H_{\rm el}}\,
\vec r\cdot\vec\nabla_R (V)\,|\phi_{\rm el}\rangle.
\nonumber
\end{eqnarray}
The function $D(R)$ depends only on the distance $R$ between the nuclei.
Although similar, $D(R)$ can not be identified with the projection of
the dipole moment operator onto the symmetry axis, because
the direction of $\vec R$ is changed under applied unitary transformation.

\section{Numerical calculations}

For the numerical calculation of $D(R)$, the clamped nuclei electronic wave 
functions were represented in the form of linear expansions in the two-electron 
basis of exponentially correlated Gaussian (ECG) functions
\begin{eqnarray}\label{EECG}
\psi_k(\br_1,\br_2)&=&\frac{1}4(1+\hat{P}_{12})(1\pm\hat\imath)\\&\times&
\exp{\left[-\sum_{i,j=1}^2 A_{k,ij}(\br_i-\bs_{k,i})(\br_j-\bs_{k,j})\right]},
\nonumber
\end{eqnarray}
where the matrices $\mathbf{A}_k$ and vectors $\bs_k$ contain nonlinear
parameters, 5 per basis function, to be variationally optimized.
The Gaussian centers $\bs_k$ were constrained to the internuclear axis
to preserve the $\Sigma$ symmetry. The antisymmetry projector
$(1+\hat{P}_{12})$ ensures singlet symmetry and the spatial projector
$(1\pm\hat\imath)$---the gerade ($+$) or ungerade ($-$) symmetry with respect to
inversion in the origin of the coordinate system located at the geometric
center of the nuclei.

The computations were performed independently at 56 internuclear distances.
In order to check the asymptotic behavior of the dipole moment function,
long distances (up to $R=12.0$ au) were sampled.
At every distance $R$, two 600-term basis sets were generated---one,
of the $^1\Sigma_g^+$ symmetry, to represent the electronic ground state wave
function $\phi_{\mathrm{el}}$, and the other, of the $^1\Sigma_u^+$ symmetry,
to invert the Hamiltonian. The parameters of the first basis set were
optimized with respect to the lowest root of the clamped nuclei Hamiltonian
$H_{\mathrm{el}}$ and the electronic energy was converged to an accuracy
of the order of a fraction of microhartree. The nonlinear parameters of the 
second basis were optimized with respect to the functional corresponding to the 
parallel polarizability
\begin{eqnarray}
\mathcal{J}=\langle \phi_{\rm el}|\vec r\cdot\vec n\,
\frac{1}{H_{\rm el}-E_{\rm el}}\,\vec r\cdot\vec n\,|\phi_{\rm el}\rangle
\end{eqnarray}
with the fixed $\phi_{\rm el}$ wave function.
The basis sets generated this way were subsequently employed to evaluate
the dipole moment $D(R)$, Eq.~(\ref{dr}). 
The proton and the deuteron mass used in Eq.~(\ref{dr})
were taken from \cite{nist}
\begin{eqnarray}
M_A &\equiv& M_{\mathrm{H}}=1836.15267247\ m_{\mathrm{e}}\\
M_B &\equiv& M_{\mathrm{D}}=3670.4829654\ m_{\mathrm{e}}.
\end{eqnarray}
The dipole moment is commonly expressed in the units of debye (D), and the 
numerical factor used in this work to convert the results from the atomic to 
debye units was 2.54174623 D/au. To inspect the saturation of the $^1\Sigma_u^+$ 
basis at $R=1.4$ au, we generated an additional 600-term basis set with the nonlinear 
parameters optimized with respect to 
\begin{equation}\label{EEF}
\langle \phi_{\rm el}|\vec r\cdot\vec\nabla_R (V)\,
\frac{1}{H_{\rm el}-E_{\rm el}}\,\vec r\cdot\vec\nabla_R (V)\,|\phi_{\rm el}\rangle
\end{equation}
and combined this basis set with the original $^1\Sigma_u^+$ basis.
Despite doubling the size of the basis set, the $D(R)$ value has changed only
on the 8-th significant figure. Hence, we expect that all
displayed figures of the final result in Table \ref{TDR} are significant. 

\begin{figure}[!ht]
\includegraphics[width=6.3cm,angle=-90]{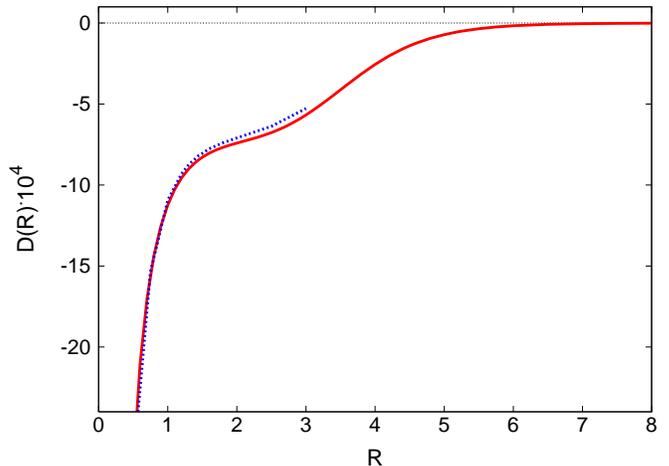}
\caption{\label{FDR} (Color online) Electric dipole moment function $D(R)$ 
(in $10^{-4}$D). Comparison to previous calculations by Ford and Browne 
\cite{ford} (dotted line). $\vec n$ is directed from deuteron to proton. 
Negative $D(R)$ means that electrons are shifted toward the deuteron.}
\end{figure}

Numerical values of the $D(R)$ function are presented in Table \ref{TDR}
\begin{table}[htb]\renewcommand{\arraystretch}{0.85}
\caption{\label{TDR} $D(R)$---the electric dipole moment (in $10^{-4}$D)
as a function of the internuclear distance $R$.
All digits are numerically significant.}
\begin{ruledtabular}
\begin{tabular*}
{0.35\textwidth}{w{3.3}w{3.4}@{\hfill}w{3.3}w{3.4}}
\cent{R/\mathrm{au}} & \cent{D(R)} &
\cent{R/\mathrm{au}} & \cent{D(R)} \\
\hline
 0.5    & -27.6224 &  3.2    & -5.0893 \\
 0.6    & -21.0635 &  3.3    & -4.7735 \\
 0.7    & -17.0294 &  3.4    & -4.4482 \\
 0.8    & -14.3747 &  3.5    & -4.1182 \\
 0.9    & -12.5426 &  3.6    & -3.7882 \\
 1.0    & -11.2339 &  3.8    & -3.1468 \\
 1.1    & -10.2754 &  4.0    & -2.5552 \\
 1.2    & -9.5603  &  4.2    & -2.0342 \\
 1.3    & -9.0193  &  4.4    & -1.5928 \\
 1.4    & -8.6054  &  4.5    & -1.4022 \\
 1.45   & -8.4353  &  4.6    & -1.2307 \\
 1.5    & -8.2853  &  4.8    & -0.9409 \\
 1.6    & -8.0347  &  5.0    & -0.7137 \\
 1.7    & -7.8352  &  5.2    & -0.5381 \\
 1.8    & -7.6721  &  5.25   & -0.5010 \\
 1.9    & -7.5336  &  5.5    & -0.3495 \\
 2.0    & -7.4102  &  5.75   & -0.2428 \\
 2.1    & -7.2933  &  6.0    & -0.1682 \\
 2.2    & -7.1758  &  6.5    & -0.0804 \\
 2.3    & -7.0517  &  7.0    & -0.0384 \\
 2.4    & -6.9156  &  7.5    & -0.0184 \\
 2.5    & -6.7632  &  8.0    & -0.0089 \\
 2.6    & -6.5913  &  8.5    & -0.0044 \\
 2.7    & -6.3972  &  9.0    & -0.0022 \\
 2.8    & -6.1799  &  9.5    & -0.0012 \\
 2.9    & -5.9390  &  10.0   & -0.0006 \\
 3.0    & -5.6755  &  11.0   & -0.0002 \\
 3.1    & -5.3911  &  12.0   & -0.0001 \\
\end{tabular*}
\end{ruledtabular}
\end{table}
and plotted in Figure~\ref{FDR}.
For comparison with previous calculations, the dipole moment function obtained
by Ford and Browne \cite{ford} in the range of $R\in \langle 0.5,3.0 \rangle$ au
is presented in the same Figure ~\ref{FDR}. The function $D(R)$ behaves as 
$R^{-2}$ at $R\to 0$, and as $R^{-4}$ at $R\to\infty$.
The singularity at $R= 0$ comes from the neglecting higher order terms
in the unitary transformation and from the adiabatic approximation.
At $R$ of the order $\sqrt{m_{\rm e}/\mu_{\rm n}}\approx 0.03$ adiabatic 
approximation fails and our formula for $D(R)$ is not valid.
At this region, however, nuclear wave function is negligible.

For the calculation of the electric dipole moments, the adiabatic potential 
of the nuclear Schr{\"o}dinger equation (\ref{07}) has been composed 
of the clamped nuclei energy $E_{\mathrm{el}}$, and the adiabatic correction 
$\langle \phi_{\rm el}|H_{\mathrm n}|\phi_{\rm el}\rangle$.
For $E_{\mathrm{el}}$ we used the energy points computed with nanohartree 
accuracy by Cencek from 1200-term ECG wave functions \cite{Cen-pc}. 
The adiabatic correction in Eq.~(\ref{Ead}) was evaluated by us using 
Eq.~(\ref{39}). The adiabatic potential curve was then obtained by means of
piecewise polynomial interpolation. The radial Schr{\"o}dinger equation
has been solved numerically using  the Le Roy's code \cite{Level}.
The obtained nuclear wave functions $\chi$ of rovibrational levels
were subsequently used in the evaluation of  the dipole
transition moments of Eq.~(\ref{18}) for the $J\to J+1$ (branch $R$)
and for the $J\to J-1$ transitions (branch $P$).

\section{Results and Discussion}

\begin{center}
\begin{table*}[!htb]
\renewcommand{\arraystretch}{1.0}
\caption{\label{ETM} Experimental and theoretical electric dipole transition
moments $D_{\rm fi}$ (in $10^{-4}$D). Relative uncertainty of our results 
due to the nonadiabatic corrections to $D_{\rm fi}$ is about $10^{-3}$.}
\begin{ruledtabular}
\begin{tabular}{l*{9}{w{4.5}}}
Reference    &\cent{P(3)}&\cent{P(2)}&\cent{P(1)}&\cent{R(0)}&\cent{R(1)}&\cent{R(2)}&\cent{R(3)}\\
\hline
\multicolumn{8}{c}{0-0}\\
experiment \cite{nelson} & & & & & 9.36(30)& 8.00(20)& 9.79(30) \\
experiment \cite{NT83} & & & & & 8.78(2)& 8.47(2)& 10.21(2) \\
theory \cite{wol} & & & & 8.36 & 8.38 & 8.39 & 8.41 \\
theory \cite{ford} &  8.282 &  8.297 &  8.306 &  8.306 &  8.297 &  8.282 &  8.262 \\
theory \cite{thorson} & 8.440 & 8.455 & 8.463 & 8.463 & 8.455 & 8.440 & 8.420 \\
this work &  8.536 &  8.551 &  8.560 &  8.560 &  8.551 &  8.536 &  8.516 \\
\\
\multicolumn{8}{c}{1-0}\\
experiment \cite{kellar1}& 0.330(40)& 0.405(30)& 0.450(30)& 0.515(20)& 0.550(30)& 0.615(30)& 0.655(40) \\
experiment \cite{RJM82}&0.340(22)&0.379(12)&0.435(11)&0.504(12)&0.533(14) & &0.609(13) \\
theory \cite{wol} & & & &  0.598 &  0.628 &  0.656 &  0.685 \\
theory \cite{ford} & 0.401 & 0.445 & 0.485 & 0.560 & 0.594 & 0.623 & 0.650 \\
theory \cite{thorson} & 0.374 & 0.421 & 0.466 & 0.552 & 0.592 & 0.630 & 0.665 \\
this work &  0.3776 &  0.4248 &  0.4708 &  0.5579 &  0.5983 &  0.6362 &  0.6714 \\
\\
\multicolumn{8}{c}{2-0}\\
experiment \cite{kellar1}& & &  0.17(2)&  0.19(2)&  0.20(2)& & \\
theory \cite{wol} & & & & 0.160 & 0.166 & 0.170 & 0.174 \\
theory \cite{ford} &  0.156 &  0.167 &  0.176 &  0.192 &  0.199 &  0.206 &  0.210 \\
theory \cite{thorson} & 0.156 & 0.167 & 0.179 & 0.200 & 0.210 & 0.219 & 0.228 \\
this work &  0.1576 &  0.1692 &  0.1805 &  0.2022 &  0.2122 &  0.2216 &  0.2301 \\
\\
\multicolumn{8}{c}{3-0}\\
experiment \cite{kellar1}& & & & 0.0795(35)& 0.0800(50)& &  \\
theory \cite{wol} & & & &  0.100 &  0.102 &  0.103 &  0.104 \\
theory \cite{ford} & 0.068 & 0.072 & 0.076 & 0.082 & 0.084 & 0.085 & 0.087 \\
theory \cite{thorson} & 0.0698 & 0.0742 & 0.0786 & 0.0870 & 0.0909 & 0.0945 & 0.0979 \\
this work &  0.0705 &  0.0749 &  0.0794 &  0.0878 &  0.0918 &  0.0955 &  0.0989 \\
\\
\multicolumn{8}{c}{4-0}\\
experiment \cite{kellar4}& & & 0.0397(26) & 0.0417(24)& 0.0425(21)& 0.0459(26)& 0.0514(53) \\
theory \cite{wol} & & & & 0.056 & 0.056 & 0.055 & 0.053 \\
theory \cite{ford} &  0.033 &  0.035 &  0.038 &  0.039 &  0.040 &  0.042 &  0.042 \\
theory \cite{thorson} & 0.0324 & 0.0345 & 0.0365 & 0.0405 & 0.0425 & 0.0442 & 0.0458 \\
this work &  0.0327 &  0.0348 &  0.0369 &  0.0409 &  0.0428 &  0.0446 &  0.0462 \\
\\
\multicolumn{8}{c}{5-0}\\
experiment \cite{dalby}& 0.0105(25)& 0.0124(21)& 0.0143(17)& 0.0181(17)& 0.0200(21)& 0.0219(25)& 0.0238(29) \\
experiment \cite{kellar4}&&&&0.0207(20)&0.0214(14)&0.0231(21)\\
theory \cite{ford} & 0.020 & 0.021 & 0.021 & 0.023 & 0.023 & 0.024 & 0.024 \\
theory \cite{thorson} & 0.0163 & 0.0173 & 0.0184 & 0.0205 & 0.0215 & 0.0225 & 0.0233 \\
this work &  0.0164 &  0.0175 &  0.0186 &  0.0207 &  0.0217 &  0.0227 &  0.0235 \\
\end{tabular}
\end{ruledtabular}
\end{table*}
\end{center}

Our electric dipole moments for the transitions between the lowest vibrational
and rotational levels are listed in Table~\ref{ETM}. Except for 0-0 transitions,
they are in good agreement with the previous calculation by Thorson {\em et al.} 
For the lowest band all theoretical predictions differ slightly from each other. 
Our results are numerically accurate to all digits shown, but the last digit is
uncertain due to the neglected $O(m_{\rm e}/\mu_{\rm n})\approx 8\times 10^{-4} $ 
higher order nonadiabatic corrections.
These corrections have been also neglected in calculations of \cite{ford,thorson},
so in principle these calculations should agree with each other.
Considering calculations presented in  \cite{wol},
we note that  in the initial expression for $\vec D_{\rm fi}$,
Wolniewicz uses $H''_{\rm n}$, Eq.~(\ref{EHbis}), as a perturbation,
and assumes the adiabatic approximation for the wave function,
but in the denominator he includes $H'_{\rm n}$ from Eq.~(\ref{EHprime}).
This expression, in comparison to our, involves
some higher order terms, namely $X_1$ of Eq.~(\ref{36}) from Appendix.
However, we show in the Appendix the cancellation
of significant contributions involving the second derivative of $\chi$
between $X_1$ and the other higher order contributions $X_2$ (Eq.~(\ref{37}))
and $X_3$ (Eq.~(\ref{38})),
which has been neglected in Wolniewicz calculations by assuming
the adiabatic wave function. Therefore, we think, slight difference with
results of Wolniewicz in \cite{wol} may come from less consistent
treatment of higher order nonadiabatic effects.

In comparison to experimental values, we observe a moderate agreement 
for all transitions, but the 0-0 ones. Here, our results, as well as the other
theoretical calculations, differ from the experiment 
by several standard deviations. We note however, that the measurements
are most cumbersome for these transitions.  As a consequence, experimental values
significantly change with the rotational number $J$ for $v=0$,
which can not be justified by theoretical analysis.
Within the ground vibronic state, the nuclear wave functions corresponding
to the lowest rotational levels are localized near the average internuclear
distance $R_0$, and differ very little from each other. For this reason
subsequent transition moments must change slowly with the rotational quantum 
number $J$, and are approximately equal to $D(R_0)$,
but the experimental results of \cite{nelson,NT83} are not consistent with 
theoretical predictions.

\section{Conclusions}

We have derived a simple expression for the electric dipole
transition rates, in terms of the dipole moment function $D(R)$, and performed 
precise calculations of $D(R)$ in a wide range of $R$. The obtained formula,
can easily be extended to other diatomic molecules, consisting of
two isotopes of the same element.  Our results for the dipole moments 
of HD molecule are numerically accurate to four digits, in moderate agreement 
with previous calculations in \cite{wol, ford, thorson} and experimental results
of \cite{nelson,NT83, bejar, kellar1,RJM82, kellar4, dalby}
(see Table \ref{ETM}). We estimate, that the relativistic and nonadiabatic 
corrections are of relative order of $10^{-4}$ and $10^{-3}$, correspondingly. 
As no other effect may alter the theoretical predictions, we suppose,
that our results are even more accurate than the experimental values obtained so far.

\section*{Acknowledgments}
We are indebted to L. Wolniewicz for valuable comments.

\section*{Appendix}
For the comparison with previous works \cite{wol, ford}, 
which used the term
\begin{equation}
H''_{\rm n} = - \frac{1}{2}\,\biggl(\frac{1}{M_B}-\frac{1}{M_A}\biggr)\,
             \vec\nabla_R\cdot\sum_a\vec\nabla_a \label{EHbis}
\end{equation}
as a perturbation, we derive the formula for $\vec D_{\rm fi}$, Eq. (\ref{17}),
using the nonadiabatic perturbation theory.
 For this, one has to abandon the assumption
in Eq.~(\ref{05}) of a separation of the electronic wave function from the 
nuclear one. Namely, the total wave function
\begin{equation}
\phi = \phi_{\rm a} + \delta\phi_{\rm na} = \phi_{\rm el}\,\chi + \delta\phi_{\rm na}
\label{26}
\end{equation}
will be the sum of the adiabatic solution and a nonadiabatic correction.
The nonadiabatic correction $\delta\phi_{\rm na}$ is  decomposed into two parts
\begin{equation}
\delta\phi_{\rm na} = \phi_{\rm el}\,\delta\chi + \delta'\phi_{\rm na},\label{27}
\end{equation}
which  obey the following orthogonality conditions
\begin{eqnarray}
\langle\delta'\phi_{\rm na}|\phi_{\rm el}\rangle_{\rm el} &=& 0\,,\label{28}\\
\langle\delta\chi|\chi\rangle &=& 0\,, \label{29}
\end{eqnarray}
with the normalization $\langle\phi|\phi_a\rangle=1$.
In the leading order of perturbative treatment, the nonadiabatic corrections
to the wave function are the following \cite{nonad}
\begin{eqnarray}
|\delta'\phi_{\rm na}^{(1)}\rangle &=&
\frac{1}{(E_{\rm el}-H_{\rm el})'}\,H_{\rm n}\,|\phi_{\rm el}\,\chi\rangle,\label{30}\\
|\delta\chi\rangle &=& \frac{1}{\bigl[E_{\rm a}- E_{\rm el}-H_{\rm n}
-\bigl\langle H_{\rm n}\bigr\rangle_{\rm el}\bigr]'}
\bigl\langle\phi_{\rm el}\bigl|H_{\rm n}\bigr|
\delta'\phi_{\rm na}\bigr\rangle_{\rm el},
\nonumber \\ \label{31}
\end{eqnarray}
where the prime in the denominator denotes subtraction of the reference
state from the Hamiltonian inversion. For the calculation of $D(R)$
one needs also the second order correction 
\begin{eqnarray}
|\delta'\phi_{\rm na}^{(2)}\rangle &=&
\frac{1}{(E_{\rm el}-H_{\rm el})'}\,H_{\rm n}\,|\phi_{\rm el}\,(\chi+\delta\chi)\rangle
+\frac{1}{(E_{\rm el}-H_{\rm el})'}
\nonumber \\&&\times
[H_{\rm n}+E_{\rm el}-E_{\rm a}]\,
\frac{1}{(E_{\rm el}-H_{\rm el})'}\,H_{\rm n}\,|\phi_{\rm el}\,\chi\rangle,
\nonumber \\ \label{32}
\end{eqnarray}
where $\delta\chi$ is given by Eq.~(\ref{31}).

The derivation of the formula (\ref{dr}) proceeds as follows.
One takes Eq. (\ref{16}) with perturbed wave functions, 
\begin{eqnarray}
\vec D_{\rm fi} &=&
\langle\phi_{\rm el}\,(\chi_{\rm f}+\delta\chi_{\rm f})+\delta'\phi_{\rm f,na}|\vec r
|\phi_{\rm el}\,
(\chi_{\rm i}+\delta\chi_{\rm i})+\delta'\phi_{\rm i,na}\rangle,\nonumber\\
&=& \vec D_{\rm fi}^{(1)}+\vec D_{\rm fi}^{(2)}\label{33}\\
\vec D_{\rm fi}^{(1)} &=&
\langle\phi_{\rm el}\,\chi_{\rm f}|\vec r\,
\frac{1}{E_{\rm el}-H_{\rm el}}\,H''_{\rm n}|\phi_{\rm el}\,\chi_{\rm i}\rangle
\nonumber \\ &&
+\langle\phi_{\rm el}\,\chi_{\rm f}|H''_{\rm n}\,
\frac{1}{E_{\rm el}-H_{\rm el}}\,\vec r|\phi_{\rm el}\,\chi_{\rm i}\rangle\,,\label{34}
\end{eqnarray}
and $\vec D_{\rm fi}^{(2)}$ is given in Eq.~(\ref{43}).
In the $\vec D_{\rm fi}^{(1)}$ one separates out the electronic matrix elements
from the nuclear ones, namely
\begin{eqnarray}
D^{(1)k}_{\rm fi} &=& -\frac{1}{2}\,\biggl(\frac{1}{M_B}-\frac{1}{M_A}\biggr)\,
\int d^3R\,\biggl\{\chi^\star_{\rm f}\,\chi_{\rm i}
\nonumber \\ &&\times
2\,\langle\phi_{\rm el}|r^k\,\frac{1}{E_{\rm el}-H_{\rm el}}\,
\vec\nabla_R \cdot \sum_a \vec\nabla_a |\phi_{\rm el}\rangle
\nonumber \\ &&
+\bigl(\chi^\star_{\rm f}\,\nabla_R^j\chi_{\rm i}
+\chi_{\rm i}\,\nabla_R^j\chi^\star_{\rm f}\bigr)
\nonumber \\ &&
\times\langle\phi_{\rm el}|r^k\,\frac{1}{E_{\rm el}-H_{\rm el}}\,
\sum_a \vec\nabla_a^j |\phi_{\rm el}\rangle\biggr\}, \label{35}
\end{eqnarray}
where the superscripts $j$ and $k$ are the Cartesian indices.
The second term in braces is integrated by parts and
$D^{(1)k}_{\rm fi}$ becomes
\begin{eqnarray}
D^{(1)k}_{\rm fi} &=& -\frac{1}{2}\,\biggl(\frac{1}{M_B}-\frac{1}{M_A}\biggr)\,
\int d^3R\,\chi^\star_{\rm f}\,\chi_{\rm i}
\nonumber \\ &&\times
\bigl[ \langle\phi_1^k|\nabla_R^j\phi_2^j\rangle-
\langle\nabla_R^j\phi_1^k|\phi_2^j\rangle\bigr], \label{36}
\end{eqnarray}
where
\begin{eqnarray}
|\phi_1^k\rangle &=& \frac{1}{E_{\rm el}-H_{\rm el}}\,r^k|\phi_{\rm  el}\rangle,
\label{37}\\
|\phi_2^j\rangle &=& \sum_a\nabla_a^j|\phi_{\rm el}\rangle\nonumber \\ &=&
-(H_{\rm el}-E_{\rm el})\,m_{\rm e}\,r^j\,|\phi_{\rm el}\rangle. \label{38}
\end{eqnarray}
One takes $\vec\nabla_R$ of the Schr\"odinger equation (\ref{06}) to obtain
\begin{eqnarray}
\nabla^j_R\phi_{\rm el} &=& \frac{1}{(E_{\rm el}-H_{\rm el})'}\,
\nabla^j_R(V)\,\phi_{\rm el}, \label{39}\\
\nabla^j_R\phi^k_1 &=&\frac{1}{E_{\rm el}-H_{\rm el}}\,
\biggl[\nabla^j_R(V-E_{\rm el})\,
\frac{1}{E_{\rm el}-H_{\rm el}}\,r^k\,\phi_{\rm el}\nonumber\\&&
+r^k\,\frac{1}{(E_{\rm el}-H_{\rm el})'}\,
\nabla^j_R(V)\,\phi_{\rm el}\biggr], \label{40} \\
\nabla^j_R\phi^j_2 &=& -\nabla^j_R(V-E_{\rm el})\,
m_{\rm e}\,r^j\,\phi_{\rm el}\nonumber \\ &&
+(E_{\rm el}-H_{\rm el})\,m_{\rm e}\,r^j\,\frac{1}{(E_{\rm el}-H_{\rm el})'}\,
\nabla^j_R(V)\,\phi_{\rm el}. \nonumber \\ \label{41}
\end{eqnarray}
The gradient of the electronic functions with respect to the internuclear
distance in Eq.~(\ref{36}) is replaced by Eqs. (\ref{40}) and (\ref{41}).
Among the four terms, two cancel out and the two other are the same, so the
transition dipole moment takes the form
\begin{eqnarray}
D^{(1)k}_{\rm fi} &=& \biggl(\frac{m_{\rm e}}{M_B}-\frac{m_{\rm e}}{M_A}\biggr)\,
\int d^3R\,\chi^\star_{\rm f}\,\chi_{\rm i}\label{42}\\ &&\times
\langle\phi_{\rm el}|r^k\,
\frac{1}{E_{\rm el}-H_{\rm el}}\,
\vec r\cdot\vec\nabla_R(V-E_{\rm el})\,|\phi_{\rm el}\rangle.
\nonumber
\end{eqnarray}
One notes that it differs from Eq.~(\ref{dr}) only by the presence of
$\vec\nabla_R E_{\rm {el}}$. We show below, that this term cancels 
out with $D^{(2)k}_{\rm fi}$ given by
\begin{eqnarray}
\vec D_{\rm fi}^{(2)} &=&
\langle\phi_{\rm el}\,\chi_{\rm f}|\vec r\,
\frac{1}{(E_{\rm el}-H_{\rm el})'}\,[H_{\rm n}+E_{\rm el}-E_{\rm i,a}]
\nonumber \\ &&\times
\frac{1}{(E_{\rm el}-H_{\rm el})'}\,H_{\rm n}\,
|\phi_{\rm el}\,\chi_{\rm i}\rangle
\nonumber \\&&+
\langle\phi_{\rm el}\,\chi_{\rm f}|\,H_{\rm n}\,
\frac{1}{(E_{\rm el}-H_{\rm el})'}\,\vec r\,
\frac{1}{(E_{\rm el}-H_{\rm el})'}\,H_{\rm n}
\nonumber \\&&\times
|\phi_{\rm el}\,\chi_{\rm i}\rangle+
\langle\phi_{\rm el}\,\chi_{\rm f}|H_{\rm n}\,
\frac{1}{(E_{\rm el}-H_{\rm el})'}
\nonumber \\ &&\times
[H_{\rm n}+E_{\rm el}-E_{\rm f,a}]
\frac{1}{(E_{\rm el}-H_{\rm el})'}\,\vec r\,
|\phi_{\rm el}\,\chi_{\rm i}\rangle.
\nonumber \\ \label{43}
\end{eqnarray}
where we neglected $\delta\chi$.
For low lying rovibrational states $\delta\chi/\chi$ is small, namely
of $O(m_{\rm e}/\mu_{\rm n})$ and thus its magnitude is of order $10^{-3}$,
if not less, and thus negligible. 
$H_{\rm n}$ is decomposed into the even $H'_{\rm n}$, Eq.~(\ref{EHprime}),
and the odd $H''_{\rm n}$ parts Eq.~(\ref{EHbis}). 
$\vec D_{\rm fi}^{(2)}$ involves the terms with a single power of $H'_{\rm n}$.
The resulting 6 terms in $\vec D_{\rm fi}^{(2)}$ we group into pairs
as follows
\begin{eqnarray}
\vec D_{\rm fi}^{(2)} &=& \vec X_1+\vec X_2+\vec X_3,\label{44}\\
\vec X_1 &=& \langle\phi_{\rm el}\,\chi_{\rm f}|\vec r\,
\frac{1}{(E_{\rm el}-H_{\rm el})}\,[H'_{\rm n}+E_{\rm el}-E_{\rm i,a}]
\nonumber \\ &\times&
\frac{1}{(E_{\rm el}-H_{\rm el})}\,H''_{\rm n}\,
|\phi_{\rm el}\,\chi_{\rm i}\rangle+\langle\phi_{\rm el}\,\chi_{\rm f}|H''_{\rm n}\,
\frac{1}{(E_{\rm el}-H_{\rm el})}\nonumber\\&\times&
[H'_{\rm n}+E_{\rm el}-E_{\rm f,a}]
\frac{1}{(E_{\rm el}-H_{\rm el})}\,\vec r\,
|\phi_{\rm el}\,\chi_{\rm i}\rangle,\label{45}\\ \nonumber \\
\vec X_2 &=& \langle\phi_{\rm el}\,\chi_{\rm f}|H'_{\rm n}\,\frac{1}{(E_{\rm  el}-H_{\rm el})'}
\,\vec r\,\frac{1}{(E_{\rm  el}-H_{\rm el})}\,H''_{\rm n}\,|\phi_{\rm el}\,\chi_{\rm i}\rangle
\nonumber\\&+&
\langle\phi_{\rm el}\,\chi_{\rm f}|H''_{\rm n}\,\frac{1}{(E_{\rm  el}-H_{\rm el})}
\,\vec r\,\frac{1}{(E_{\rm  el}-H_{\rm el})'}\,H'_{\rm n}\,|\phi_{\rm el}\,\chi_{\rm i}\rangle,
\nonumber \\ \label{46} \\
\vec X_3 &=& \langle\phi_{\rm el}\,\chi_{\rm f}|\vec r\,\frac{1}{(E_{\rm  el}-H_{\rm el})}
\,H''_{\rm n}\,\frac{1}{(E_{\rm  el}-H_{\rm el})'}\,H'_{\rm n}\,|\phi_{\rm el}\,\chi_{\rm i}\rangle
\nonumber\\&+&
\langle\phi_{\rm el}\,\chi_{\rm f}|H'_{\rm n}\,\frac{1}{(E_{\rm  el}-H_{\rm el})'}
\,H''_{\rm n}\,\frac{1}{(E_{\rm  el}-H_{\rm el})}\,\vec r\,|\phi_{\rm el}\,\chi_{\rm i}\rangle.
\nonumber \\\label{47}
\end{eqnarray}
$H'_{\rm n}$ involves two derivatives over $R$ and $H''_{\rm n}$ a single
derivative. Consider terms with three derivatives of $\chi$.
Since $\chi$ satisfies Eq.~(\ref{07}), the second derivative of $\chi$
coming from $H'_{\rm n}$ cancels with $E_{\rm a} - E_{\rm el}(R)$,
leaving a small term $\langle H'_{\rm n}\rangle_{\rm el}$
and the derivative of $E_{\rm el}$. Thus
no term with three derivatives of $\chi$ is present. We will show below
that  no terms involving any derivatives of $\chi$ are present.
Each $\vec X_i$ includes two derivatives terms
\begin{eqnarray}
X_1^k &\approx& \frac{1}{2\,\mu_n}\,\biggl(\frac{1}{M_B}-\frac{1}{M_A}\biggr)\,
\int d^3R\,\bigl[\chi^*_{\rm f}\,(\nabla_R^l\,\nabla_R^j\,\chi_{\rm i})
\nonumber\\&&
+\chi_{\rm i}\,(\nabla_R^l\,\nabla_R^j\,\chi^*_{\rm f})\bigr]\,
\langle\phi_{\rm el}|r^k\,\frac{1}{E_{\rm el}-H_{\rm el}}\,r^j|
\nabla_R^l\phi_{\rm el}\rangle,\nonumber \\ \label{48}\\
X_2^k &\approx& \frac{1}{2\,\mu_n}\,\biggl(\frac{1}{M_B}-\frac{1}{M_A}\biggr)\,
\int d^3R\,\bigl[(\nabla_R^l\chi^*_{\rm f})\,(\nabla_R^j\,\chi_{\rm i})\nonumber\\ &&+
(\nabla_R^j\chi^*_{\rm f})\,(\nabla_R^l\,\chi_{\rm i})\bigr]\,
\langle\phi_{\rm el}|r^j r^k\,\frac{1}{E_{\rm el}-H_{\rm el}}|
\nabla_R^l\phi_{\rm el}\rangle,\nonumber \\ \label{49}\\
X_3^k &\approx& -X_1^k-X_2^k\,,\label{50}
\end{eqnarray}
but all of them cancel out in the sum.
A single derivative of $\chi$ has to be of the form
$\vec\nabla_R (\chi^*_{\rm f}\,\chi_{\rm i})$, what can be integrated by parts
and transformed into derivatives of $\phi_{\rm el}$ and the resolvent.
Considering terms without derivatives of $\chi$ all of them are of order
$O[(m_{\rm e}/m_{\rm n})^2]$ but one, which involves $\vec\nabla_R E_{\rm el}$
which arised from the commutator 
$[H'_{\rm n}+E_{\rm el}(R)-E_{\rm a},\vec\nabla_R]$ in $X_1$ in Eq.~(46). 
It is of the form
\begin{equation}
\vec D_{\rm fi}^{(2)} = \biggl(\frac{m_{\rm e}}{M_B}-\frac{m_{\rm e}}{M_A}\biggr)\,
\langle\phi_{\rm el}\chi_{\rm f}|\vec r\frac{1}{(E_{\rm el}-H_{\rm el})'}
\vec r\cdot\vec\nabla_R\,(E_{\rm el})|\phi_{\rm el}\chi_{\rm i}\rangle
\label{51}
\end{equation}
which together with $\vec D_{\rm fi}^{(1)}$ gives the leading
correction to the transition dipole moment Eq.~(\ref{17}).

\end{document}